# PARAMETRIC PROBABILISTIC APPROACH FOR CUMULATIVE FATIGUE DAMAGE USING DOUBLE LINEAR DAMAGE RULE CONSIDERING LIMITED DATA

João Paulo Dias [a], Stephen Ekwaro-Osire [a,*], Americo Cunha Jr. [b], Shweta Dabetwar [a], Abraham Nispel [a], Fisseha M. Alemayehu [c], Haileyesus B. Endeshaw [d]

[a] Texas Tech University, Department of Mechanical Engineering, 805 Boston Avenue, Lubbock, TX 79409, USA

[b] Rio de Janeiro State University, Institute of Mathematics and Statistics, 524 São Francisco Xavier Street, Rio de Janeiro, RJ 20550-900, Brazil

[c] West Texas A&M University, School of Engineering, Computer Science and Mathematics, 2501 4th Avenue, Canyon, TX 79015, USA

[d] Colorado State University, Department of Mechanical Engineering, 400 Isotope Dr, Fort Collins, CO 80521, USA

[*]Corresponding author: phone +1 (806) 834-1308, e-mail stephen.ekwaro-osire@ttu.edu

**Abstract:** This work proposes a parametric probabilistic approach to model damage accumulation using the double linear damage rule (DLDR) considering the existence of limited experimental fatigue data. A probabilistic version of DLDR is developed in which the joint distribution of the knee-point coordinates is obtained as a function of the joint distribution of the DLDR model input parameters. Considering information extracted from experiments containing a limited number of data points, an uncertainty quantification framework based on the Maximum Entropy Principle and Monte Carlo simulations is proposed to determine the distribution of fatigue life. The proposed approach is validated using fatigue life experiments available in the literature.

**Keywords:** uncertainty quantification, cumulative fatigue damage, double linear damage rule, Maximum Entropy Principle, limited data experiments.

1. Introduction

Structural damage due to fatigue is considered one of the major issues in the reliability of engineering structures subjected to cyclic loads regimes. Fatigue damage increases with the applied loading cycles in a cumulative manner and the prediction of fatigue life is a crucial step in preliminary design in order to avoid unexpected failure of critical mechanical components [1,2]. Although several cumulative fatigue damage (CFD) models have been proposed in the past decades, none of them have wide acceptance and more research is needed in order to develop sufficiently general CFD models for reliable life prediction of engineering structures [3]. Existing CFD models are grouped into six major categories [4,5]: linear damage rules, non-linear and double linear damage approaches, life curve modification methods, approaches based on crack growth concepts, continuum damage mechanics models and energy-based theories. Among these models, the linear damage rule (LDR) model is the oldest and most widely used in engineering applications due to its simplicity. It was first proposed by Palmgren [6] in 1924 and reintroduced in its classical version later by Miner [7]. However, it has been recognized that LDR is unresponsive to load-level sequence and uncommon cumulative damage, for example, when the metallurgical effect occurs at high-temperature loadings [8]. These limitations have been observed and alternative models have been proposed to overcome the issues with LDR. One conservative solution was proposed by Marco and Starkey [9] as a non-linear version of the LDR to improve its drawbacks. In between the linear and non-linear damage approaches, the double linear damage rule (DLDR), proposed by Manson [10,11], was developed to overcome the deficiencies of the LDRs associated with the loading sequence. The key concept behind the DLDR involves the simplification of the non-linear model using two straight lines connected by a knee-point, in which each line was originally associated to the physical processes of crack initiation and propagation [10]. However, this association was later abandoned by their own authors, referring to the straight lines simply as phase I and phase II [11]. Many authors [5,12–14] have justified the application of DLDR to solve non-linear damage accumulation problems under multi-amplitude loading conditions based on the accurate predictions obtained with a relatively simple mathematical formulation since it requires the determination of only one knee-point. More sophisticated generalized non-linear CFD models have been proposed recently [15], which are computationally demanding and more difficult to be implemented in a probabilistic context.Although most of the above-mentioned CFD approaches are essentially deterministic models, experimental studies have been shown a considerable scattering of the fatigue life for a wide range of materials and loading conditions, revealing the stochastic nature of the cumulative fatigue damage [16]. Therefore, probabilistic approaches should be considered to carefully account for the various sources of uncertainties present in the CFD model parameters. Some probabilistic approaches have been presented considering mainly the LDR and non-linear versions of the LDR. Rathod et al. [17] proposed a non-stationary linear fatigue damage accumulation model combined with a probabilistic S–N curve method applied to multi-stress loading regimes. Pinto et al. [16] used the Palmgren-Miner rule to determine the cumulative distribution function of fatigue life of components submitted to three load levels assuming Weibull and lognormal distributions. Sun et al. [4] proposed a CFD model based on the Palmgren-Miner rule to calculate the statistical characteristics of fatigue life under variable amplitude loading conditions. Using a Weibull S–N field model originally proposed by Castillo and Fernandez-Canteli [18], Blason et al. [19] presented a probabilistic CFD approach based on Miner-Palmgren rule. Recently, Zhu et al. [20] proposed a CFD model under random loadings based on the combination of a probabilistic version of the LDR combined with finite element analysis for high-pressure turbine discs. Liu and Mahadevan [21] proposed a methodology which combines a non-linear version of the LDR and a stochastic S–N curve representation technique for fatigue life prediction under variable loadings. Following a similar approach, Zhu et al. [22] proposed an approach based on a non-linear damage accumulation concept and a probabilistic S-N curve to model damage accumulation of railway axle steels. Acknowledging the contribution of these probabilistic CFD works, they still carry the shortcomings associated with the LDR (which may provide inaccurate predictions for multi-load regimes) and non-linear versions of the LDR (which may be computationally expensive). Correia et al. [5] proposed the only known probabilistic CFD approaches based on the DLDR and the Weibull S–N field model proposed by Castillo and Fernandez-Canteli [18].

The uncertainty characterization techniques proposed in most of the above-mentioned works are based on the collection of data from fatigue experiments, in which uncertainties of the model parameters were described using traditional statistical parametric regression methods assuming Weibull and lognormal distributions. However, the determination of statistically significant fatigue life data by experiments is very expensive and time-consuming, and when limited experimental data is available, traditional regression methods are difficult to apply [23,24]. Entropy methods, such as the Maximum Entropy Principle (MaxEnt), are viable alternatives to model the distribution of fatigue life by reducing subjective uncertainty from the introduction of assumed distribution types when limited or no experimental data are available [25,26]. Aiming to address the gaps highlighted above, this work proposes a parametric probabilistic CFD approach to quantify the uncertainties of the DLDR model parameters considering the existence of limited experimental data. A probabilistic version of the DLDR for the two-loading levels was developed in which the joint probability distribution of the coordinates knee-point was obtained as a function of the probability distributions of the DLDR model input parameters. Based on statistical information extracted from existing experimental datasets with a limited number of samples, an uncertainty quantification framework based on the MaxEnt Principle and on Monte Carlo simulations was proposed to determine the probability distributions of the coordinates of the knee-point and the fatigue life. The proposed probabilistic DLDR approach was validated using fatigue life data for two-load level experiments available in the literature. Furthermore, results obtained with the classic LDR and a recently proposed single-parameter non-linear model were implemented in the proposed probabilistic framework and compared to the DLDR predictions.

## 2. Methodology

### 2.1. Deterministic Modeling

A schematic illustration of deterministic approaches of linear, non-linear and double linear models for two-level loading sequence is presented in Figure 1. In this figure, fatigue life is described by the relationship between the applied cycle ratio, $\beta_1 = n_1/N_1$, and the remaining cycle ratio $\beta_2 = n_2/N_2$, where $n_1$ and $n_2$ are the number of cycles applied for the first load and the number of remaining cycles to failure when the second load is applied, respectively. $N_1$ and $N_2$ are, respectively, the fatigue lives when the first and second load levels are individually applied. In the linear model (LDR), frequently referred to as the Palmgren-Miner rule, the cumulative damage, $D$, is defined by the linear summation of the applied and remaining cycle ratios, $\beta_1$ and $\beta_2$, as [7],

$$D = \beta_1 + \beta_2 = \frac{n_1}{N_1} + \frac{n_2}{N_2}, \tag{1}$$

where for a given number of cycles applied to the first load level, $n_1$, the number of remaining cycles, $n_2$, is automatically determined considering that the component fails when $D$ approaches unity. The non-linear model is based on an exponential relationship between the applied and remaining life cycles, in which the exponent is usually a material parameter dependent of the lead level [9]. Among several existing non-linear models, Rege and Pavlou [27] proposed a one-parameter model based on iso-damage curves converging at the knee-point of the S-N curve of the material. For a two-level loading sequence, the remaining cycles to failure is described in terms of the following logarithm relationship,

$$\log(N_2 - n_2) = \log N_e - \frac{\log N_e - \log N_2}{\left(\frac{\log N_e - \log N_1}{\log N_e - \log n_1}\right)^{\frac{q(\sigma_1)}{q(\sigma_2)}}}, \tag{2}$$

where $N_e$ is the number of cycles related to the endurance limit of the material and the exponent $q(\sigma_1)/q(\sigma_2)$ is the model parameter, which is a function of the cyclic stress levels, $\sigma_1$ and $\sigma_2$. The model parameter can be determined by using two-load level fatigue experiments. On the other hand, the double linear model (DLDR) approximates the non-linear model using two straight lines, which divides the fatigue life into Phase I and Phase II, connected by a knee-point. The DLDR only requires the definition of the knee-point which, in terms of the coordinates in the axis $\beta_1$ and $\beta_2$, is defined as,

$$\left.\frac{n_1}{N_1}\right|_{knee} = \beta_{1,knee} = (1 - B)\left(\frac{N_1}{N_2}\right)^{\alpha}, \tag{3}$$

$$\left.\frac{n_2}{N_2}\right|_{knee} = \beta_{2,knee} = B\left(\frac{N_1}{N_2}\right)^{\alpha}, \tag{4}$$

where $\alpha$ and $B$ are two DLDR parameters that need to be determined experimentally. The procedure to obtain the parameters $\alpha$ and $B$ from experiments is detailed in [10,11].

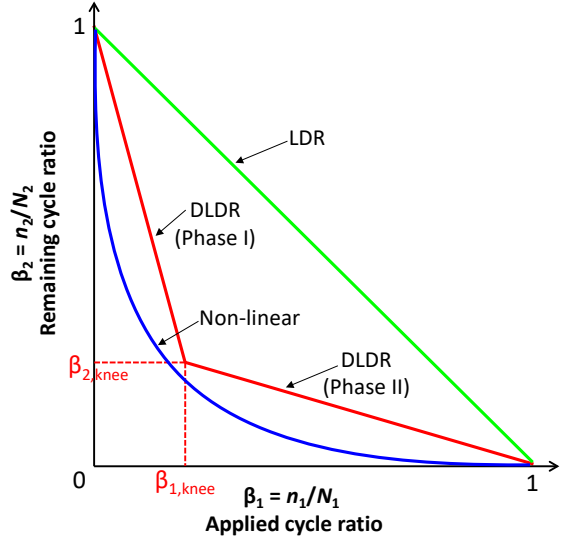

**Figure 1.** Deterministic approach for the linear (LDR), non-linear, and double linear (DLDR) fatigue damage models.

A further step towards the deterministic formulation of the DLDR can be made by describing the relationship between $\beta_1$ and $\beta_2$ for phases I and II, respectively, through algebraic functions of the coordinates of the knee-point $\beta_{1,knee}$ and $\beta_{2,knee}$ as,

$$\beta_2 = \left(\frac{\beta_{2,knee} - 1}{\beta_{1,knee}}\right)\beta_1 + 1 \quad \text{for} \quad 0 \leq \beta_1 \leq \beta_{1,knee}, \tag{5}$$

$$\beta_2 = \beta_{2,knee}\left(1 + \frac{\beta_{1,knee}}{1 - \beta_{1,knee}}\right)(1 - \beta_1) \quad \text{for} \quad \beta_{1,knee} \leq \beta_1 \leq 1. \tag{6}$$

Given the primitive relationships $\beta_1 = n_1/N_1$ and $\beta_2 = n_2/N_2$, Eqs. (5) and (6) can be rearranged explicitly in terms of $n_1$ and $n_2$ as,

$$n_2 = \left[\left(\frac{\beta_{2,knee} - 1}{\beta_{1,knee}}\right)\frac{n_1}{N1} + 1\right]N_2 \quad \text{for} \quad 0 \leq \frac{n_1}{N_1} \leq \beta_{1,knee}, \tag{7}$$

$$n_2 = \beta_{2,\text{knee}} \left(1 + \frac{\beta_{1,\text{knee}}}{1 - \beta_{1,\text{knee}}}\right)\left(1 - \frac{n_1}{N_1}\right) N_2 \quad \text{for} \quad \beta_{1,\text{knee}} \leq \frac{n_1}{N_1} \leq 1. \tag{8}$$

Equations (7) and (8) are convenient ways to describe the remaining fatigue life, $n_2$, as functions of all other DLDR parameters. Notice that in these equations, information about $N_1$, $N_2$, α, and $B$ are implicit in the coordinates of the knee-point – see Eqs. (3) and (4). This feature will be explored in detail in the probabilistic approach of the DLDR.

## 2.2. Probabilistic Double Linear Damage Rule

### 2.2.1. Overview of the Proposed Probabilistic DLDR Approach

Before going through the details of the proposed probabilistic DLDR framework, it is worthwhile to present an overview of a conventional deterministic DLDR approach compared with the proposed probabilistic DLDR approach for two-load levels applications. Figure 2 depicts an overall comparison between the two approaches. In conventional deterministic approaches, deterministic quantities for each DLDR input parameters are defined and the model equations, Eqs. (3), (4), (7), and (8), are solved deterministically in order to obtain the output parameters. For the proposed probabilistic approach, first, probability distributions of the input random variables need to be defined using uncertainty modeling techniques. Then the uncertainties of the input parameters need to be propagated through the model equations and finally, statistical inference is used to obtain the probability distributions of the DLDR output parameters.

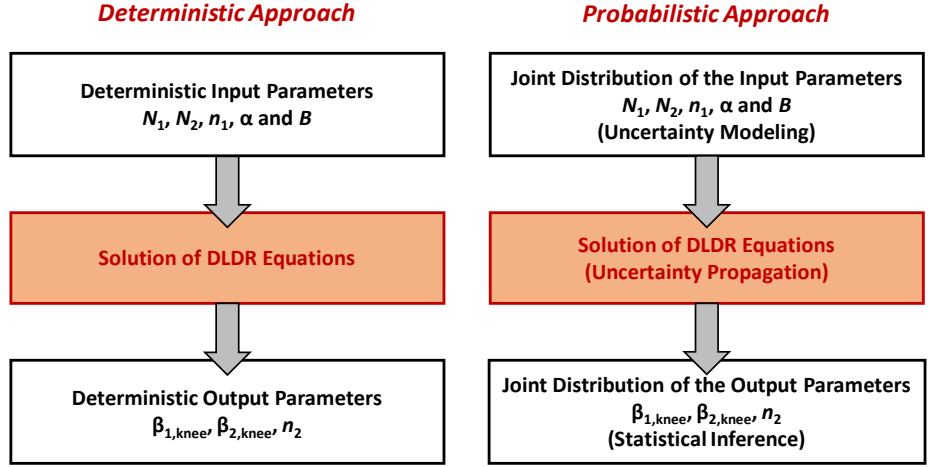

**Figure 2.** Comparison between conventional deterministic and the probabilistic DLDR approaches.

Unlike the deterministic approach, the coordinates of the knee-point in the proposed probabilistic DLDR are defined through a joint probability density function (joint PDF), $f_{\beta1,\text{knee},\beta2,\text{knee}}$, as graphically shown in Figure 3. In other words, the joint PDF provides information about the location of the knee-point coordinates for specific probability levels.

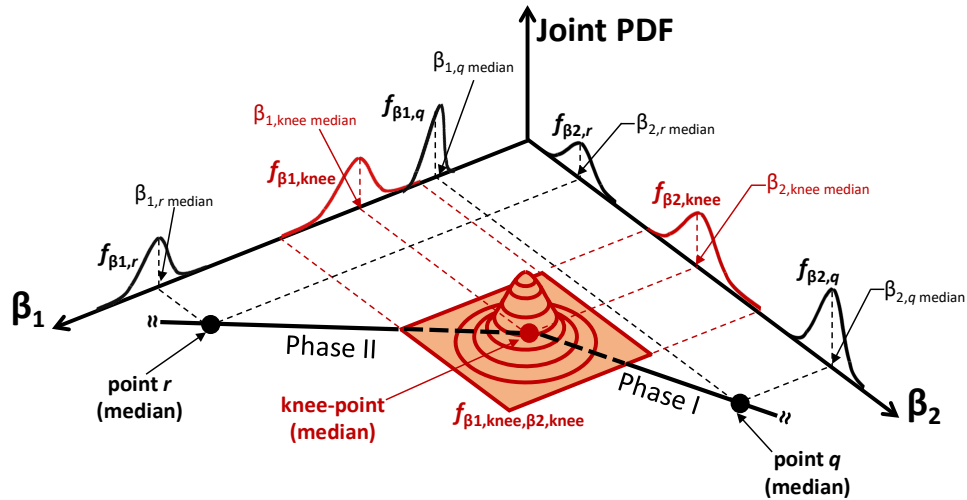

**Figure 3.** Graphical interpretation of the probabilistic DLDR.

For instance, if the probability level of the 50[th] percentile (median) is of interest, then $\beta_{1,\text{knee median}}$ and $\beta_{2,\text{knee median}}$ can be obtained from the marginal PDFs of $\beta_{1,\text{knee}}$ and $\beta_{2,\text{knee}}$, denoted as $f_{\beta1,\text{knee}}$ and $f_{\beta2,\text{knee}}$, respectively, (see the red dot in Figure 3). Although the knee-point is a singular point that defines

Phase I and Phase II of the DLDR (see Figure 1), the same rationale is applicable to any point belonging to these phases. Figure 3 also shows the generic points *q* and *r* in phases I and II respectively, which are obtained from the median of the marginal PDFs of their respective coordinates.

The joint PDF $f_{\beta_{1,knee},\beta_{2,knee}}$ can also be used to determine the conditional probability that the knee-point is in a specific area of the DLDR graph. This concept is illustrated in Figure 4. In this figure, the LDR is used as a reference since $\beta_{1,knee} + \beta_{2,knee} = 1$ in this line, which divides the DLDR graph into the high-low load sequence area, where $\beta_{1,knee} + \beta_{2,knee} < 1$ and the low-high sequence area, where $\beta_{1,knee} + \beta_{2,knee} > 1$. If one is interested in determining the conditional PDF for the knee-point located in the high-low load sequence area, then for a fixed value for $\beta_{1,knee} = \beta_{1,knee\ median}$ and considering that the coordinates of the knee-point are statistically dependent random variables, the conditional PDF is written as [28],

$$f_{\beta_{2,knee}|\beta_{1,knee}} = \frac{f_{\beta_{1,knee},\beta_{2,knee}}}{f_{\beta_{1,knee}}}. \tag{9}$$

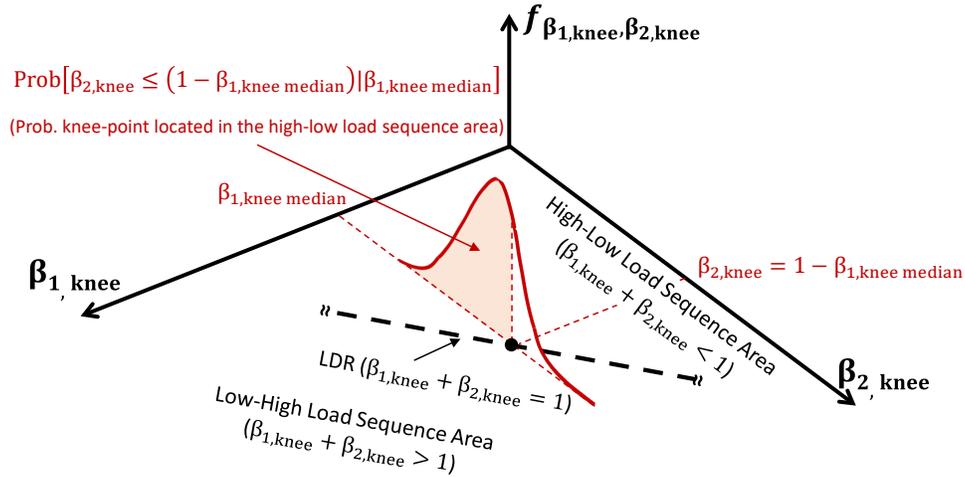

**Figure 4.** Schematic representation of the probabilistic position of the knee-point and the loading sequence in DLDR.

If $\beta_{1,\text{knee}}$ is fixed as the median, then the conditional probability that the knee-point is located in the high-low load sequence area can be determined by integrating the conditional PDF given by Eq. (9) as shown in Figure 4 to obtain,

$$\text{Prob}_{\text{H-L area}} = \text{Prob}\big[\beta_{2,\text{knee}} \leq (1 - \beta_{1,\text{knee median}})|\beta_{1,\text{knee median}}\big]. \tag{10}$$

2.2.2. Uncertainty Modeling

The first step in the specification of a probabilistic model for the DDLR is the construction of the joint PDF for the model input parameters, since, as discussed above, it is necessary to provide a statistical characterization of the response. The construction of a consistent probabilistic model for the joint PDF must be done based only on known information about the parameters in order to avoid bias introduced by presumed information. In this sense, the joint PDF must be constructed based on a rational criterion and can never be arbitrated. Two different scenarios are considered: (i) sufficient experimental data is available; (ii) few or no experimental data is available.

In the first scenario, the rational approach employs non-parametric statistics (no algebraic form for the joint PDF is assumed) to infer the joint PDF of parameters. Non-parametric estimators such as empirical cumulative distribution function (empirical CDF), and kernel density estimator (KDE) are employed in this scenario [29]. In the second scenario, which is more frequent, a conservative approach to constructing the probabilistic model using MaxEnt is used [30,31]. This tool from information theory says that the distribution to be chosen for the vector of random parameters is the one that, in addition to being consistent with known information, maximizes the entropy. Mathematically, we look for a joint PDF that maximizes the entropy function,

$$S = -\int f_X(x)\ln f_X(x)dx, \tag{11}$$

where $f_X(x)$ is the joint PDF of the random vector $X$ composed by the DLDR input parameters, and respect the $M + 1$ constraints defined by the known information about $X$,

$$\int_{\mathbb{R}} g_k(x) p_X(x) dx = m_k, \quad \text{for} \quad k = 0, 1, \dots, M, \tag{12}$$

where $g_k(x)$ and $m_k$ are known real functions and values (generally statistical moments), respectively, with $g_0(x) = 1$ and $m_0 = 1$. For instance, for a single random variable, if the known information is the support $[a, b]$, the mean, $\mu_X$, and the second-order moment $\mu_X^2 + \sigma_X^2$, where $\sigma_X$ is the standard deviation of $X$, the PDF given by MaxEnt is,

$$f_X(x) = 1_{[a,b]}(x) \exp(-\lambda_0 - \lambda_1 x - \lambda_2 x^2), \tag{13}$$

where the parameters $\lambda_0$, $\lambda_1$, and $\lambda_2$ are the Lagrange multipliers and depends on the known statistical information about $X$, i.e., $[a, b]$, $\mu_X$ and $\sigma_X$. It is also worth mentioning that, when no cross-moment information is provided, the MaxEnt formalism provides a joint distribution that is the product of the marginal distributions of the parameters, i.e. they are independent.

2.2.3. Uncertainty Propagation

Once the probability distribution of the model input is defined, it is necessary to calculate how these uncertainties are modified by the model to give rise to the output distribution. The solution to this problem is what is called the propagation of uncertainties. The stochastic solver employed in this work to solve this uncertainty propagation problem is the Monte Carlo method, whose procedure is illustrated schematically in Figure 5. This procedure is divided into three steps, namely, pre-processing, processing and post-processing. In the pre-processing, samples are generated according to the joint PDF of the input parameters using the inverse transform method [28]. In the processing step, the model equations are solved for each of these samples, giving rise to a set of output samples, which are used in a non-parametric statistical inference process to estimate the output probability distributions in the post-processing step. This can be used to statistically characterize quantities of interest generated by the model.

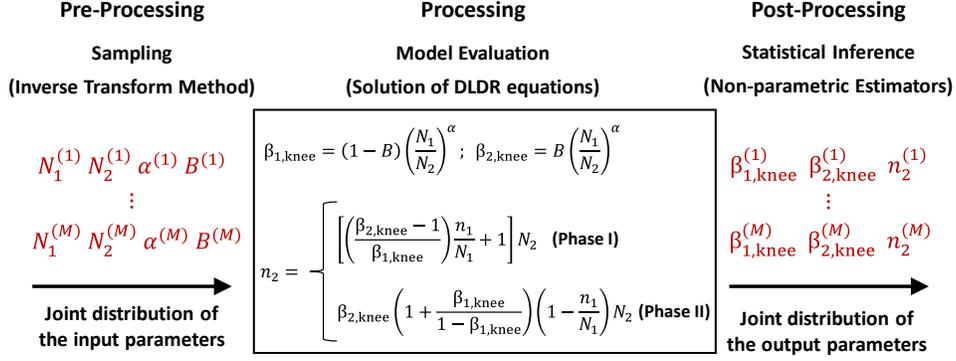

**Figure 5.** Uncertainty propagation of the DLDR input parameters using the Monte Carlo method.

It is also important mentioning that the proposed framework involves the propagation of uncertainties of four inputs parameters ($N_1$, $N_2$, $\alpha$ and $B$) to three output parameters ($\beta_{1,\text{knee}}$, $\beta_{2,\text{knee}}$, and $n_2$), which implies that an existing correlation structure is embedded in the image of input distribution by the model operator. Considering a scenario in which a lot of information (data) about all the input parameters is available, non-parametric inference methods can be employed to estimate marginal PDFs to determine the joint PDF of the input parameters and its correlation structure. However, considering that limited information about the input parameters is generally available (as will be shown later in the Results and Discussion section), a more conservative and consistent way of specifying their PDFs is through the MaxEnt principle. In this formalism, the most uncertain (thus, least biased) distribution consistent with the known information about the random parameters is specified, which provides a safe and robust criterium to choose a prior probabilistic law for the model input. Additionally, if no information on the cross statistical moments is provided, the joint PDF of the input parameters is obtained through the product of the marginal PDFs, i.e., the input the parameters are specified as being independent. Since information about covariances of the input parameters is difficult to obtain, an input random vector with independent parameters was considered. To arbitrate these covariances is a biased procedure, which can lead to an inconsistent model.

For this reason, the proposed framework does not consider any correlation structure for the input parameters.

2.2.4. Validation of the Proposed Model

The validation of the proposed probabilistic model was carried out considering experimental fatigue life datasets available in the literature from two different sources. The first source, obtained by Tanaka et al. [32] (hereafter referred to as Tanaka), has been widely used in previous cumulative fatigue works due to the substantial amount of the sample tested. The second source was obtained by Xie [33] in which smaller sample size were tested for two different steel grades: 0.45% carbon steel and 16Mn steel alloy, hereafter referred to as Xie045 and Xie16Mn, respectively. The test conditions and the total number of specimens tested for each dataset are presented in Table 1.

Table 1. Experimental datasets used to validate the proposed probabilistic model.

| Material | Stress Level or Loading Sequence [MPa] | Sample Size | # of Cycles Applied in the First Load Level ($n_1$) | Reference |
|---|---|---|---|---|
| Nickel-Silver Alloy | 478 (single-load) | 200 | N/A | Tanaka [32] |
| | 666 (single-load) | 200 | N/A | |
| | 666 → 478 (H-L) | 200 | 13,300; 26,500; 39,800 and 55,400 | |
| 0.45% Carbon Steel | 331 (single-load) | 18 | N/A | Xie045 [33] |
| | 309 (single-load) | 16 | N/A | |
| | 331 → 309 (H-L) | 38 | 40,300; 80,600 and 120,900 | |
| 16 Mn Steel Alloy | 373 (single-load) | 15 | N/A | Xie16Mn [33] |
| | 394 (single-load) | 15 | N/A | |
| | 394 → 373 (H-L) | 30 | 26,000; 44,000 and 75,000 | |

The three datasets were built for two-load levels high-low (H-L) loading sequence fatigue tests. In Tanaka, the two-load tests were divided into four groups, in which four different fixed numbers of cycles were applied in the first load level ($n_1$ equal to 13,300, 26,500, 39,800, and 55,400 cycles) and 50 specimens were tested in each group, totalizing 200 specimens. In the two-load tests conducted in Xie045, two groups containing 13 specimens and one group containing 12 specimens (totalizing 38 specimens) were considered, in which the number of cycles applied in the first load level was 40,300; 80,600 and 120,900 cycles, respectively. For Xie16Mn dataset, three batches containing 10 specimens each were tested with 26,000; 44,000 and 75,000 cycles applied in the first load level. The results of the two-load fatigue tests conducted by Tanaka and Xie were used to extract the information needed to model the uncertainties of the DLDR input variables $\alpha$ and $B$. Moreover, both authors carried out independent single-load fatigue tests for each one of the two stress levels involved in the two-load fatigue tests. The results of the single-load fatigue tests were used to model the uncertainties of the DLDR input parameters $N_1$ and $N_2$.

The flowchart of the proposed uncertainty quantification framework is shown in Figure 6. According to this flowchart, the uncertainty modeling of the DLDR input parameters is conditioned to the availability of data from the experiments of Tanaka, Xie045, and Xie16Mn for each input parameter. If a significant amount of experimental data is available, KDE was used to determine the distribution that generates the experimental data. On the other hand, if few or no experimental data is available, MaxEnt was used to estimate the distribution using the available information about the DLDR input parameter. In the case where only a few experimental points are available, the information about the statistical moments (mean and standard deviation) of the data was considered. For the special case where experimental data is unavailable, information from other sources was used to obtain the MaxEnt distributions of the DLDR input parameters.

It is also important to mention that the scenarios involving experimental datasets with only a few points may present an additional difficulty to consider the information about the statistical estimators, due to the weak statistical significance of small datasets collected from fatigue experiments. In order to overcome this difficulty, a criterion based on the mean-square convergence of the statistical estimators of a random variable is proposed to determine if the information contained in the dataset is a good representation of the statistics population [29]. Once the distributions of the DLDR input parameters are properly characterized, the inverse transform method is used to obtain the samples of the parameters. Next, the Monte Carlo simulations are carried out to propagate the uncertainties of the input parameters through the DLDR model equations to the output parameters. Finally, non-parametric statistical estimators are used to compute the statistics which in turn to obtain the distribution of the DLDR output parameters. Both KDE and empirical CDF non-parametric estimators are used for this purpose.

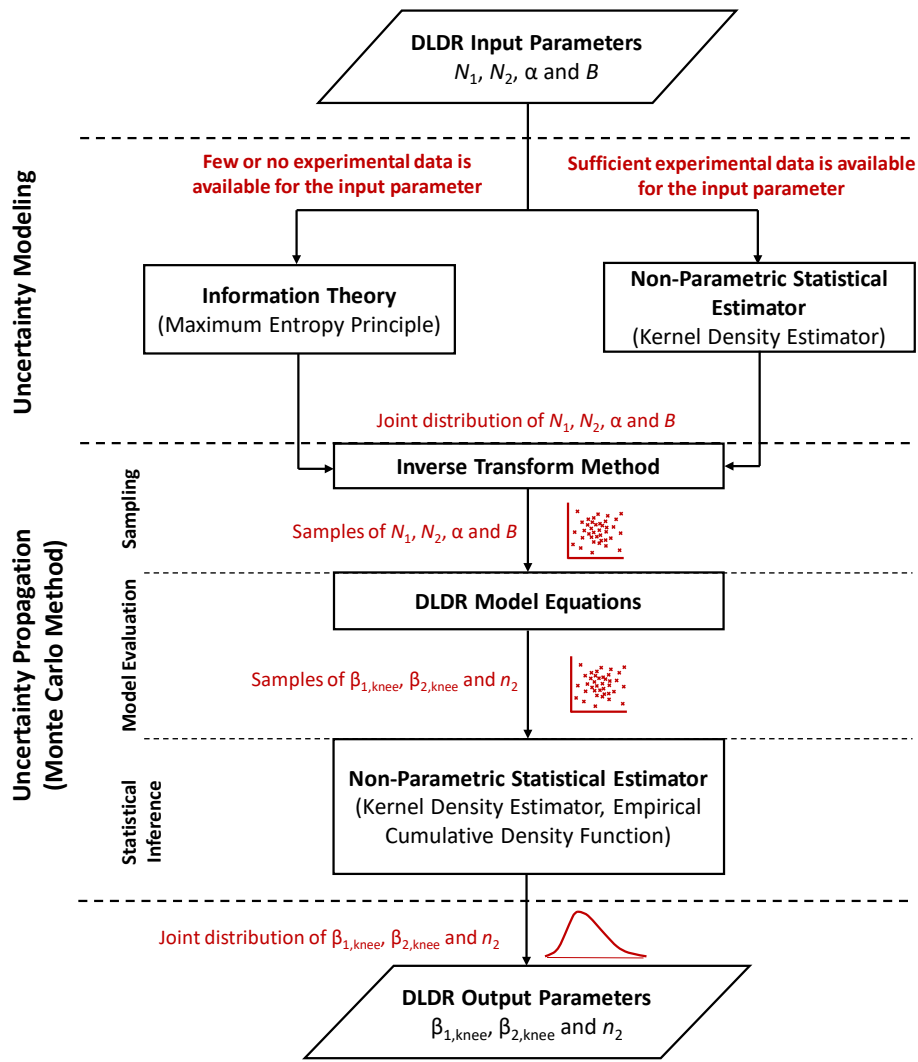

**Figure 6.** Proposed uncertainty quantification framework for the probabilistic DLDR considering the scenarios when large and/or limited data is available.

## 3. Results and Discussions

### 3.1. Dataset Convergence and Uncertainty Modeling of $N_1$ and $N_2$

Since fatigue experimental datasets with different numbers of samples were considered in this work, it is worthwhile first to determine if such datasets are a good representative of the population in order to obtain accurate sample statistics, e.g. mean, standard deviation, and PDF. For any random variable for which a sufficiently large and diverse dataset is available, non-parametric statistical estimators, such as KDE, provide the better estimations for the underlying probability distribution. A widely used criterion to ensure convergence of the distribution estimator is the mean-square criterium, which is based on the convergence of the standard deviation estimator. The convergence of the former is ensured by the convergence of the latter [29]. Figure 7 illustrates the convergence tendencies for the standard deviation estimators of $N_1$ and $N_2$ for the single-load fatigue experiments presented in Tanaka. Since the convergence history is sensitive to the arrangement of the dataset, the experimental datasets for $N_1$ and $N_2$ from Tanaka were randomly sorted three times in order to check accurately their convergence. Given the relatively large number of samples available in these datasets, it is apparent that the fluctuations in the standard deviation estimators reduces significantly as the number of samples increase. Nevertheless, the sample estimators tend to stabilize when the number of samples is increased to a certain threshold, which is roughly around 150-180 samples for the mean and standard deviation of $N_1$ and $N_2$, respectively.

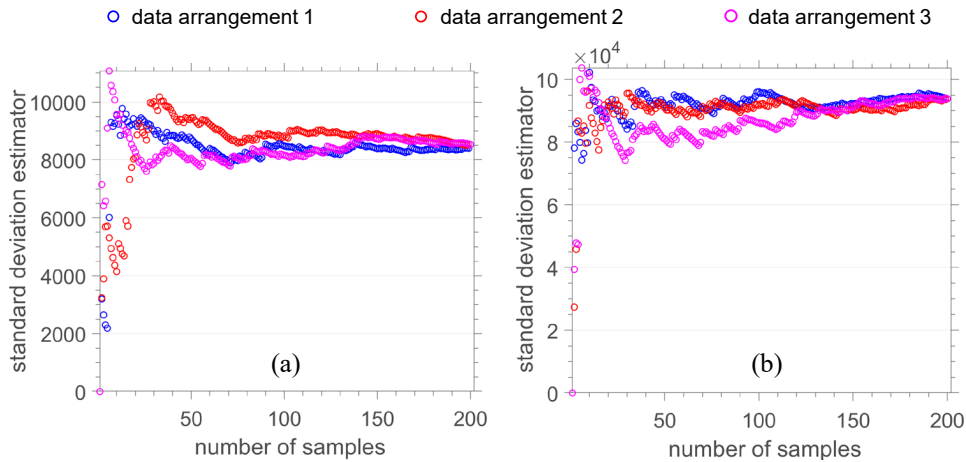

**Figure 7.** Convergence of the standard deviation of Tanaka dataset using the mean-square criterion for (a) $N_1$ and (b) $N_2$.

The mean-square criterion does not only establish the representativeness of the dataset but also analyses the convergence of the dataset in the probability distribution, thereby allowing the utilization of non-parametric approaches for representing the probability density function that generates the population data. The failure to ensure the convergence implies that the information contained in the dataset is not representative of the population, and hence the non-parametric distribution determined out of the data is biased and it may significantly vary when the dataset size changes. In order to illustrate this concept, Figure 8 presents KDE probability distributions obtained with all 200 samples for $N_1$ and $N_2$ contained in Tanaka dataset and with only 20 samples randomly selected from each dataset. Moreover, the MaxEnt probability distributions were also obtained for the same 200 and 20 randomly selected samples in order to contrast with the KDE distributions. For the distributions obtained with the MaxEnt principle, the mean and coefficient of variation (COV) were directly extracted from the datasets with 200 and 20 samples, and support considered to vary from 0 to +∞. The results show that the KDE distribution for the 200 samples differs considerably from KDE distribution obtained with the 20 samples dataset. This indicates that the lack of convergence in the statistics for $N_1$ and $N_2$ datasets with only 20 samples (see Figure 7) yields to unreliable estimations of the KDE distributions for those reduced datasets. On the other hand, since MaxEnt is a more conservative approach, which presents with very low sensitivity to the number of samples in the dataset, estimations of MaxEnt distributions with only 20 samples as accurate as KDE distributions with 200 samples were obtained. This is because MaxEnt aims to establish the probability density function with the largest level of uncertainty based on the available information at the moment of the analysis; conversely, the outcomes are the most conservative in terms of the all possible random scenarios, and thus avoiding potential overestimations of the response.

The results in Figure 8 also show that when the datasets with 200 samples are considered, the KDE and the MaxEnt provide distributions that are very close to each other. On the other hand, for the datasets with 20 samples, the discrepancy between the KDE and MaxEnt distributions is clearly observed, and this comparison can be used to indirectly estimate the convergence of the dataset. Such comparison is useful for situations where only a few experimental points were collected, and the mean-square criterion is difficult to apply. This was the scenario given by Xie045 and Xie16Mn datasets for $N_1$ and $N_2$ and the comparison between the probability distributions obtained with KDE and MaxEnt for these datasets are shown in Figure 9. A similar discrepancy can be observed as in the reduced Tanaka dataset, which confirms that the KDE cannot be applied to model the uncertainties of $N_1$ and $N_2$ using those experimental datasets with limited data. It is also worth mentioning that, due to the lack of convergence of these datasets, the KDE distributions changed drastically as the datasets were randomly rearranged. Therefore, in these limited data scenarios, MaxEnt provides a viable option to model the uncertainties of $N_1$ and $N_2$ based on the information obtained from those datasets.

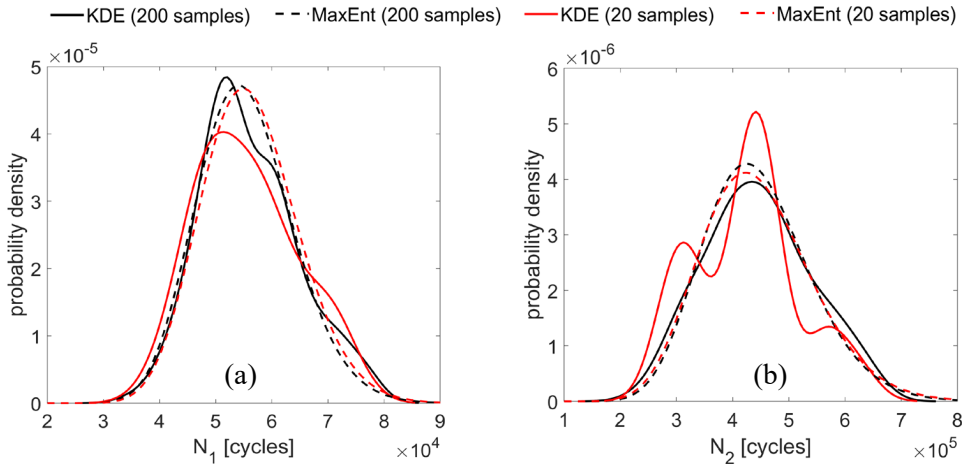

**Figure 8.** KDE and MaxEnt distributions for Tanaka datasets considering different number of samples: (a) $N_1$ and (b) $N_2$.

*3.2. Uncertainty Modeling of the DLDR Parameters α and B*

The results of the convergence analysis presented in the previous section showed that the number of samples presented in Tanaka dataset for $N_1$ and $N_2$ is sufficiently large to allow their uncertainty modeling using the non-parametric KDE. On the other hand, the application of MaxEnt to model uncertainties of the same parameters for Xie045 and Xie16Mn datasets was justified by their limited amount of samples. For the other DLDR parameters $\alpha$ and $B$, the availability of data obtained from these experimental sources was even more reduced. For this reason, the MaxEnt principle was used to estimate the probability densities of $\alpha$ and $B$ based on the pieces of information extracted indirectly from the experimental datasets. Table 2 lists the information considered for the application of the MaxEnt principle to model the uncertainties of the DLDR parameters $\alpha$ and $B$ for the datasets reported in Tanaka [32] and Xie [33].

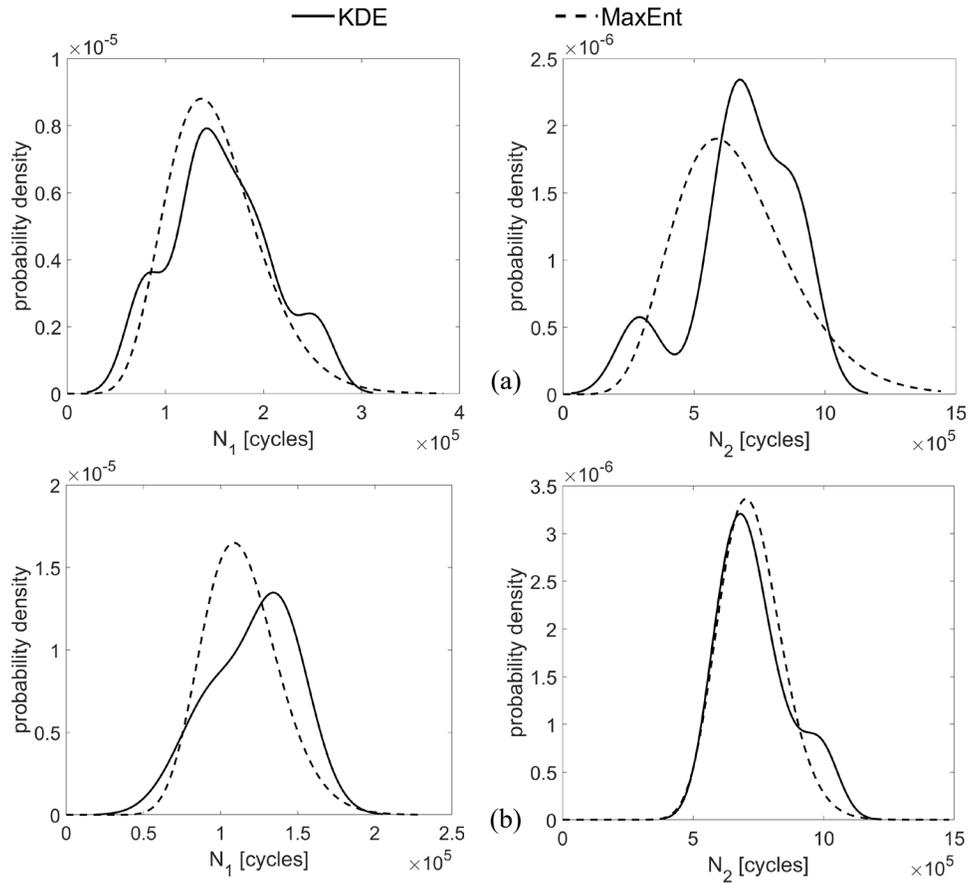

**Figure 9.** KDE and MaxEnt distributions for $N_1$ and (b) $N_2$ considering different number of samples: (a) Xie045 dataset and (b)Xie16Mn dataset.

The mean estimates of $\alpha$ and $B$ were obtained by fitting a deterministic double linear curve to the mean values of the fatigue lives of the two-load experiments. The COVs, for lack of better information available, were also considered uniform random variables within the support limits listed in Table 2. The justification for this choice is that the estimation of the dispersion of the parameters α and B is very difficult to obtain from the two-load experiments since prior-knowledge about the variability of the coordinates of the knee-point is not available.

**Table 2.** Information for the uncertainty modeling of the DLDR parameters α and B using MaxEnt.

| Dataset | Parameter | |
|---|---|---|
| | α | B |
| Tanaka [32] | Support: [-1, 1]<br>Mean: -0.03<br>COV: uniform distribution between 0.05 and 0.10 | Support: [0, 1]<br>Mean: 0.80<br>COV: uniform distribution between 0.05 and 0.10 |
| Xie045 [33] | Support: [0, 1]<br>Mean: 0.34<br>COV: uniform distribution between 0.05 and 0.10 | Support: [0, 1]<br>Mean: 0.45<br>COV: uniform distribution between 0.05 and 0.10 |
| Xie16Mn [33] | Support: [0, 1]<br>Mean: 0.50<br>COV: uniform distribution between 0.05 and 0.10 | Support: [0, 1]<br>Mean: 0.50<br>COV: uniform distribution between 0.05 and 0.10 |

Figures 10 (a) and (b) show the MaxEnt PDFs of the input parameters α and B obtained using the Tanaka and Xie045 two-load fatigue datasets, respectively. The results obtained for Xie16Mn were similar and are not shown in this section for brevity. It can be observed that the distributions of the parameters α and B present a very similar behavior regarding the curve shape. It is clear that the expected value and the dispersion of these parameters vary for different materials, and that the deterministic generic values proposed by Manson et al. [11], particularly α=0.25, may not be applicable for the cases studied here. For Tanaka's dataset, α value close to zero were found, which results in a damage curve close to the LDR.

### 3.3. Joint Probability Densities of the Coordinates of the Knee-Points

According to the framework depicted in Figure 6, once the uncertainties of the input parameters were modeled and propagated using a Monte Carlo method, distributions of the DLDR output parameters of interest were obtained. Figure 11 shows the results obtained for the joint probability densities of the coordinates of the knee-point, $\beta_{1,knee}$ and $\beta_{2,knee}$, considering Tanaka and Xie045 datasets. Once again, the results for Xie16Mn are not shown for the sake of brevity.

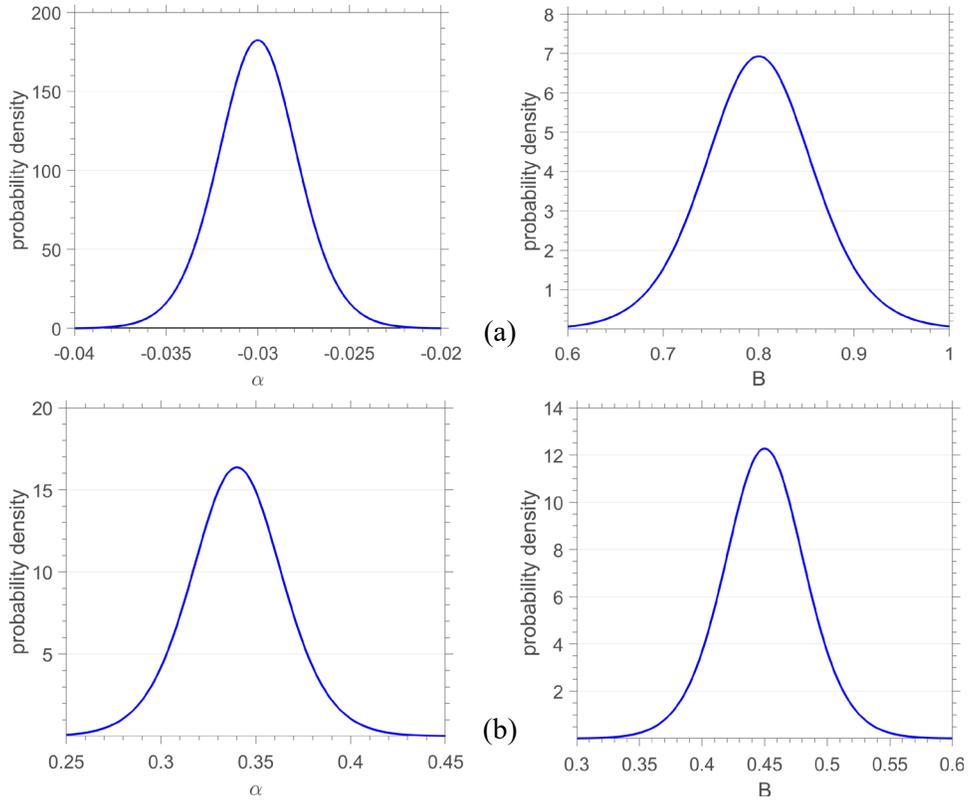

**Figure 10.** Estimated MaxEnt distributions of the DLDR input variables $\alpha$ and $B$ using the two-load experimental dataset obtained from: (a) Tanaka, and (b) Xie045.

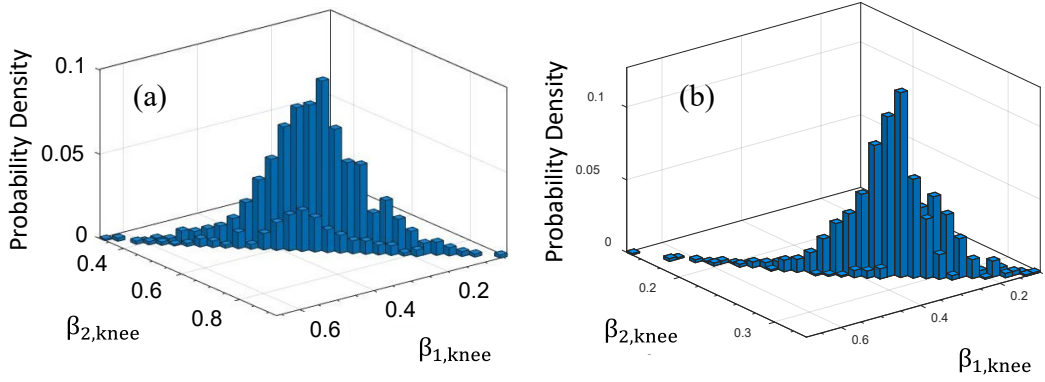

**Figure 11.** Results for the joint probability densities of $\beta_{1,knee}$ and $\beta_{2,knee}$ obtained using experimental data from (a) Tanaka, and (b) Xie045.

Given the statistical dependency between $\beta_{1,knee}$ and $\beta_{2,knee}$ shown in Eqs. (3) and (4), the joint probability densities depicted in Figure 11 can be used to determine the location of the knee-point for different probability levels. Some of these probabilities, given fixed values of $\beta_{1,knee}$, were calculated considering the LDR as a reference and the corresponding results are shown in Table 3. The results presented in Table 3 indicate that for each value of $\beta_{1,knee}$ considered, there is a probability of 99.9% that the knee-point is located in the high-low loading sequence area, i.e. bellow the LDR line, for Xie045 dataset. Due to the proximity with the linear behavior, the calculated probabilities that the knee-point is located in the high-low loading sequence area for Tanaka dataset is slightly smaller than that of Xie045 dataset for $\beta_{1,knee}$ equal to 0.50 and 0.75. It is important to mention that in the results presented in Table 3, the LDR, which divides the graphical representation of DLDR into two areas, was used as a reference. However, any location in the DLDR graph can be used as a reference and hence the probability of the knee-point being in that location can be determined. Since the location of the knee-point affects the relationship between the applied cycle ratio and the remaining cycle ratio (see Figure 1), the proposed probabilistic approach can be used to provide a rigorous description of the remaining fatigue life considering uncertainties in all DLDR model parameters.

**Table 3.** Conditional probabilities calculated from the joint distributions of $\beta_{1,knee}$ and $\beta_{2,knee}$

| $\beta_{1,knee}$ | $\text{Prob}[\beta_{2,knee} \leq (1 - \beta_{1,knee})|\beta_{1,knee}]$ | |
|---|---|---|
| | **Tanaka et al. [32]** | **Xie045 [33]** |
| 0.25 | 99.99% | 99.99% |
| 0.50 | 99.10% | 99.99% |
| 0.75 | 97.39% | 99.99% |

### 3.4. Estimation of the Remaining Fatigue Life

In the proposed probabilistic DLDR framework, explicit equations for the fatigue remaining life, $n_2$, were obtained in terms of the coordinates of the knee-point, $\beta_{1,knee}$ and $\beta_{2,knee}$, the fatigue lives for each load level, $N_1$ and $N_2$, and the number of cycles applied in the first load level, $n_1$. These relationships, as mathematically expressed in Eqs. (7) and (8), allow the calculation of the distributions of $n_2$ as a response function, which can be directly compared with the two-load experimental results. In fact, a key characteristic of the proposed probabilistic framework is that it can be easily implemented to other CFD models. In order to demonstrate such flexibility, Figure 12 - Figure 14 show the DLDR predictions for $n_2$ compared with the H-L fatigue experimental datasets and the predictions obtained with the classical linear model (LDR) and the one parameter non-linear model based on iso-damage curves proposed by Hege and Pavlou [27]. Referring to the proposed probabilistic framework in Figure 6, the random input parameters are $N_1$ and $N_2$, the model equation is given by Eq. (1), and the only random output parameter is $n_2$ for the LDR model. For the non-linear model, the random input parameters are $N_1$, $N_2$, and the function $q(\sigma_1)/q(\sigma_2)$, the model equation is given by Eq. (2), and $n_2$ is the random output parameter. The number of cycles related to the endurance limit of the material ($N_e$) was assumed to be constant and calculated from the S-N curve for each material. Similar to the DLDR parameters α and $B$, the statistical information about the function $q(\sigma_1)/q(\sigma_2)$ of the non-linear model was extracted from the two-load fatigue experimental datasets and its uncertainty was modeled using the MaxEnt principle with the COV modeled as a uniformly distributed random variable with support between 0.05 and 0.10.

Figure 12 presents the results for the predictions of the scattering of the remaining fatigue life for Tanaka's two-load level experiments, for three different values of $n_1$. From this experimental dataset, the mean value of non-linear model function $q(\sigma_1)/q(\sigma_2)$ was 2.00 and the support was [1.60, 2.60]. The results show that our probabilistic framework predicted satisfactorily the scattering of $n_2$ from Tanaka experiments for all three CFD models considered. The non-linear model provided slightly better predictions of the probability distribution, at least when $n_1$ was equal to 13,300 and 39,800 cycles (Figure 12a and c). In general terms, the small difference between the predictions of the three models can be attributed to the almost linear relationship between cycle ratios $n_1/N_1$ and $n_2/N_2$ observed in Tanaka's two-load level experiments, which may be related to the material characteristics of the Ni-Ag alloy.

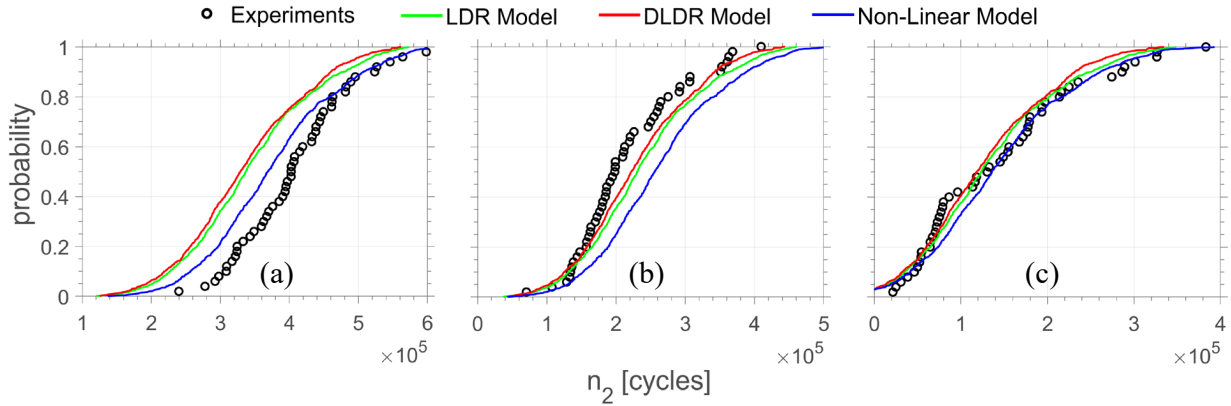

**Figure 12.** Experimental remaining fatigue life from Tanaka H-L experiments and the probability distributions predicted with the DLDR, LDR, and non-linear models: (a) *n₁*=13,300 cycles; (c) *n₁*=26,500 cycles, and (d) *n₁*=39,800.

The probability distributions of the remaining fatigue life predictions for the carbon steel of Xie045 two-load level experiments are presented in Figure 13. For this experimental dataset, the mean value of the function $q(\sigma_1)/q(\sigma_2)$ and its support were estimated as 0.63 and [0.51, 0.76], respectively. Unlike the results for Tanaka's experiments, our probabilistic framework with the DLDR model clearly provided better predictions for the scattering of $n_2$ for all values of $n_1$.

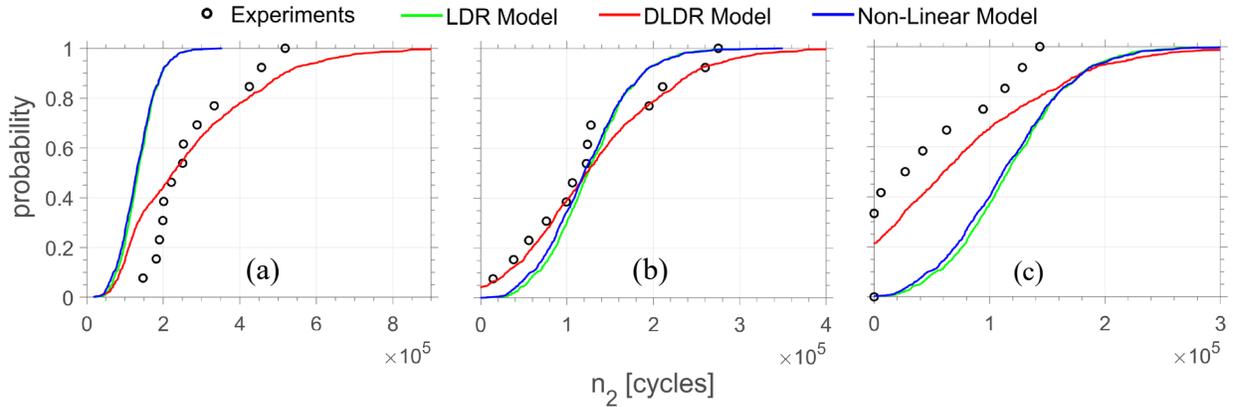

**Figure 13.** Experimental remaining fatigue life from Xie045 H-L experiments and probability distributions predicted with DLDR, LDR and non-linear models: (a) $n_1$=40,300 cycles; (b) $n_1$=80,600 cycles, and (c) $n_1$=120,900 cycles.

The results in Figure 14, for the remaining fatigue life predictions for the steel alloy from Xie16Mn two-load level experiments, confirm the same trend of the DLDR model providing the best predictions of the scattering of $n_2$ compared to the LDR and non-linear models. From the Xie16Mn experimental dataset, the estimated mean value of the function of the non-linear model was 1.74, whereas the its support was [1.50, 1.98]. In fact, the damage accumulation mechanisms of most steel grades subjected to cyclic loads present a non-linear behavior, which explains why the LDR model failed to predict the remaining fatigue life of Xie045 and Xie16Mn experimental datasets. Furthermore, the non-linear model has only one parameter ($q(\sigma_1)/q(\sigma_2)$), whereas the DLDR model has two parameters ($\alpha$ and $B$), which gives the latter model more flexibility to be adjusted to the two-load level experimental fatigue data when compared to the former model.

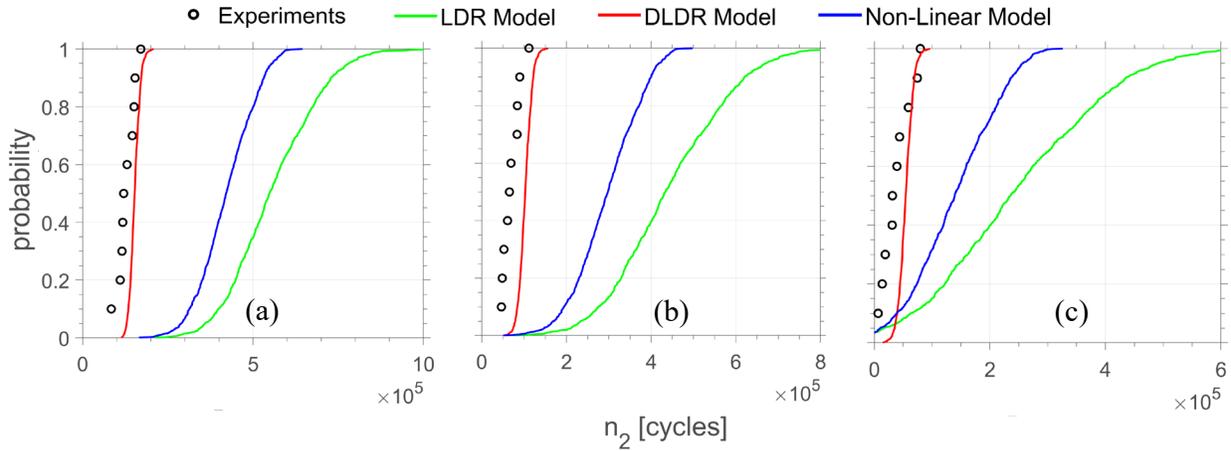

**Figure 14.** Experimental remaining fatigue life from Xie16Mn H-L experiments and probability distributions predicted with DLDR, LDR and non-linear models: (a) $n_1$=26,000 cycles; (c) $n_1$=44,000 cycles, and (d) $n_1$=75,000 cycles.

Figure 15 shows the prediction of the probability distributions of the remaining fatigue life, $n_2$ for any given value of applied cycles in the first load level, $n_1$ using Tanaka dataset. In this figure, the predicted median and 98% confidence bounds for $n_2$ are shown which were calculated from the predicted probability distributions of $n_2$ at different fixed values of $n_1$.

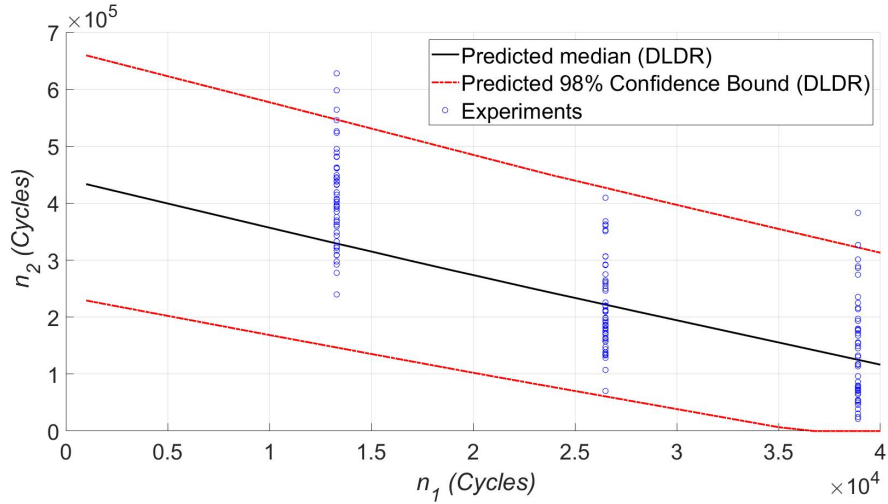

**Figure 15.** Probabilistic $n_2$ vs. $n_1$ using DLDR for the Tanaka dataset.

As it can be seen, the fatigue life curve is almost linear as the mean value of α is very small. Moreover, most of the experimental data points representing the variability of $n_2$ for three different values of $n_1$ falls within the predicted 98% confidence bounds, which demonstrates the accuracy of the proposed model to predict the variability of the remaining fatigue life of two-load level experiments. Figure 16 shows the values of the median prediction curve with the 98% percent confidence bounds for $n_2$ for fixed values of $n_1$ using Xie045 datasets. From the graph, it can be clearly seen that the datasets for three different load levels fall inside of the confidence bounds. Additionally, it can be observed that the predicted median line approximately matches the central tendency of the experimental datasets, at least for the $n_1$ equal to 80,600 and 120,900. In contrast with Figure 15, the current figure presents an inflection point when $n_1$ is approximately $4 \times 10^4$ cycles, which is explained by the differences in the parameter α between the Tanaka and Xie045 datasets.

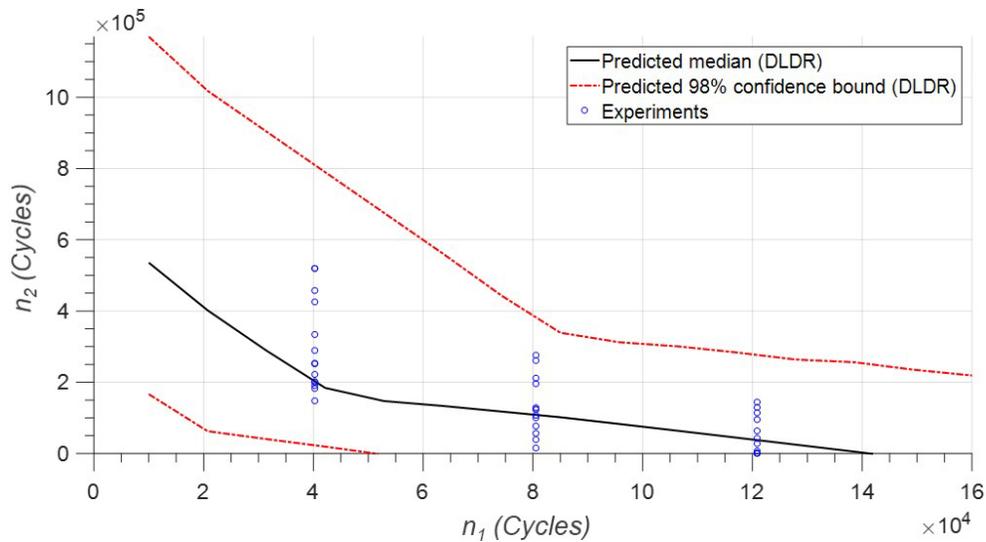

**Figure 16.** Probabilistic $n_2$ vs. $n_1$ using DLDR for the Xie045 data.

*3.5. Characteristics of the Proposed Framework*

In this paper, we propose a rigorous approach to quantify uncertainties of the DLDR model parameters based on the availability of fatigue experimental data for each parameter. The framework was systematically divided into two essential phases, namely, uncertainty modeling and uncertainty propagation phases. The key strength of our method compared to other probabilistic CFD approaches, such as the probabilistic DLDR method proposed by Correia et al. [5], relies on the fact that regardless of the amount of data/information available for the model parameters, we avoid bias in assuming Weibull or lognormal distributions to model their uncertainties. Although these classic parametric approaches have been widely used in literature, they simply fit curves to the data, not providing any guarantees that these distributions generate the observed data. Non-parametric estimators such as KDE provide the most unbiased probability distribution models for scenarios in which enough fatigue experimental data is available for a specific random parameter. When limited data is available, on the other hand, parametric uncertainty modeling approaches based on curve fitting from the data become even more questionable, since the data is not statistically representative, as we showed in Xie's fatigue experiments. For this scenario, MaxEnt is a viable alternative to provide the most unbiased probability distribution model for the CFD model parameters.

Despite its advantages, the probabilistic framework also carries shortcomings related to the uncertainty modeling methods. First, as shown in section 3.1, the application of non-parametric methods, e.g. kernel density estimator, requires a considerable amount of fatigue data, which is not always possible to obtain experimentally. Furthermore, even for the most common scenario containing limited fatigue data, the accuracy of the probability distribution estimated with the MaxEnt is sensitive to the quality of the information available. If poor quality information is available, or if poor quality experiments were carried out, it may compromise the quality of the information and lead to poor estimations of the uncertainties of a random variable using MaxEnt.

## 4. Conclusions

This paper presented a novel probabilistic interpretation of the double linear damage rule (DLDR). A rigorous uncertainty quantification approach based on non-parametric statistics, the Maximum Entropy Principle, and Monte Carlo simulation was proposed taking into consideration the availability of experimental data to model the uncertainties of the DLDR input parameters. The model was used to predict the variability of the fatigue life of two-load high-low sequence experiments from the literature. The conclusions are:

- The mean-square convergence criterion applied for the standard deviation estimators for the DLDR input parameters $N_1$ and $N_2$ showed that Tanaka dataset, with 200 samples for each load level, is more statistically significant than Xie045 and Xie16Mn datasets, each one containing less than 20 samples for $N_1$ and $N_2$. For this reason, the kernel density estimator method was used to model the uncertainties of $N_1$ and $N_2$ for the Tanaka's dataset, whereas the Maximum Entropy principle was used to provide the most conservative estimation of the uncertainties of $N_1$ and $N_2$ for Xie045 and Xie16Mn datasets.

- Due to insufficient data available for the DLDR input parameters $\alpha$ and $B$ for both Tanaka, Xie045 and Xie16Mn datasets, the Maximum Entropy principle was also used to provide a conservative estimation of the uncertainties of these parameters considering the mean value estimated from the two-load fatigue experiments and the coefficient of variation as a uniform random variable with known support limits.

- A novel probabilistic interpretation of the DLDR taking into consideration the statistical dependency between the coordinates of the knee-points was provided, in which the location of the knee-points can be determined for different probability levels.

- Although the proposed probabilistic framework was originally developed considering the DLDR model, it was demonstrated that it can be easily incorporated to other CFD models. Comparisons with the framework implemented with the linear model and a one-parameter non-linear model showed that the DLDR model best represented the scattering of the remaining fatigue life for the two-load level experiments carried out on two different steel grades. For the experiments on Ni-Ag alloy performed by Tanaka, no considerable difference in results were observed among the three CFD models.

- Future contributions for this research may include the extension of the proposed approach to other classes of materials and more complex loading regimes, and the realization of sensitivity analysis to determine which input parameters have more impact on the model response.